\documentclass[aps,prev,preprintnumbers,floatfix,nofootinbib,twocolumn]{revtex4-1}

\pdfoutput=1

\usepackage{mathrsfs, amsmath, amsthm, amssymb, color, epstopdf, verbatim, enumerate, graphicx, hyperref}

\begin{document}

\title{Searching for solar axions using data from the Sudbury Neutrino Observatory}
\author{Aagaman Bhusal$\,{}^{a,b}$}
\author{Nick Houston$\,{}^{c}$}\email{nhouston@bjut.edu.cn}
\thanks{corresponding author.}
\author{Tianjun Li$\,{}^{a,b}$}
\affiliation{${}^a$ CAS Key Laboratory of Theoretical Physics, Institute of Theoretical Physics, Chinese Academy of Sciences, Beijing 100190, China\\
		${}^b$ School of Physical Sciences, University of Chinese Academy of Sciences, No. 19A Yuquan Road, Beijing 100049, China\\
		${}^c$ Institute of Theoretical Physics, Faculty of Science, Beijing University of Technology, Beijing 100124, China}

\begin{abstract}
We explore a novel detection possibility for solar axions, which relies only on their couplings to nucleons, via the axion-induced dissociation of deuterons into their constituent neutrons and protons. 
An opportune target for this process is the now-concluded Sudbury Neutrino Observatory (SNO) experiment, which relied upon large quantities of heavy water to resolve the solar neutrino problem.
From the full SNO dataset we exclude in a model-independent fashion isovector axion-nucleon couplings $|g^3_{aN}|\equiv\frac{1}{2}|g_{an}-g_{ap}|>2\times10^{-5}$GeV${}^{-1}$ at 95 \% C.L. for sub-MeV axion masses, covering previously unexplored regions of the axion parameter space.
In the absence of a precise cancellation between $g_{an}$ and $g_{ap}$ this result also exceeds comparable constraints from other laboratory experiments, and excludes regions of the parameter space for which astrophysical constraints from SN1987A and neutron star cooling are inapplicable due to axion trapping.
\end{abstract}

% keywords: solar, axion, axion like particle nuclear, nucleon, deuterium, SNO

\maketitle

\textbf{Introduction.} 
Arising straightforwardly as a minimal extension of the Standard Model, and in particular the Peccei-Quinn solution of the strong CP problem \cite{Peccei:1977ur, Weinberg:1977ma,Wilczek:1977pj}, axions and axion-like particles (ALPs) occupy a rare focal point in theoretical physics, in that they are also simultaneously a generic prediction of the exotic physics of string and M-theory compactifications \cite{Svrcek:2006yi, Arvanitaki:2009fg}.
Despite the profound differences between these contexts the resulting axion properties are largely universal, creating an easily-characterisable theoretical target.

As typically light, long lived pseudoscalar particles they can also influence many aspects of cosmology and astrophysics, leading to a wealth of observational signatures \cite{Marsh:2015xka}.
In particular, they provide a natural candidate for the mysterious dark matter comprising much of the mass of our visible universe \cite{Dine:1982ah, Preskill:1982cy}, and as such are an focal point of intense ongoing investigation \cite{Irastorza:2018dyq}.

Fortuitously, our own Sun should provide an intense and readily available axion flux from which much of the corresponding parameter space can be constrained.
At present the strongest resulting constraint is provided by the CAST experiment \cite{Anastassopoulos:2017ftl}, which relies upon the Primakoff conversion of axions into photons in a background magnetic field.
Presumably for reasons of observational and experimental convenience, most of the existing axion literature rests similarly on axion-photon interactions. 

We will in the following instead turn attention to a less well-explored corner of the parameter space, and in particular the axion-nucleon interactions
\begin{equation}
	\mathcal{L}=\frac{1}{2}g_{an}\left(\partial_\mu a\right)\,\overline n\gamma^\mu\gamma_5n+\frac{1}{2}g_{ap}\left(\partial_\mu a\right)\,\overline p\gamma^\mu\gamma_5 p\,,
\end{equation}
which can be re-expressed in terms of the neutron/proton doublet $N = (n,p)$ as
\begin{equation}
	\mathcal{L}=\frac{1}{2}\left(\partial_\mu a\right)\overline N\gamma^\mu\gamma_5\left(g^1_{aN}I+g^3_{aN}\tau_3\right)N\,,
	\label{derivative action}
\end{equation}
where the isoscalar and isovector couplings are respectively $g^1_{aN}=(g_{an}+g_{ap})/2$, $g^3_{AN}=(g_{an}-g_{ap})/2$, and $\tau_3=$diag(1,-1).
Although we will not make use of this here, if required integration by parts can be used to recast this into
\begin{equation}
	\mathcal{L}=-ia\overline N\gamma_5m_N\left( g^1_{aN}I+ g^3_{aN}\tau_3\right)N\,, 
\end{equation}
where $m_N=\text{diag}(m_n, m_p)$.
We also emphasise here that although we consider axion/nucleon interactions, we are not assuming a QCD axion specifically.

There are comparatively few constraints on axion-nucleon interactions, arising primarily from considerations of SN1987A \cite{Engel:1990zd,Chang:2018rso,Lee:2018lcj,Carenza:2019pxu}
and neutron star (NS) observations \cite{Sedrakian:2015krq, Sedrakian:2018kdm, Beznogov:2018fda, Hamaguchi:2018oqw}, experimental searches for new spin-dependent forces \cite{Fischbach:1999bx, Masso:1999hi, Vasilakis:2008yn, Bezerra:2014dja, Bezerra:2014ona, Mostepanenko:2016pdc,Adelberger:2006dh} and time dependent nuclear electric dipole moments \cite{Abel:2017rtm}.
It should however be noted that the overall paradigm of axion constraints derived from SN1987A has been called into question \cite{Blum:2016afe, Bar:2019ifz}.

The CASPER experimental program is also searching for axion dark matter via nuclear interactions \cite{Budker:2013hfa, Wu:2019exd, Garcon:2019inh}, and similar considerations have led recently to novel constraints from existing comagnetometer data \cite{Bloch:2019lcy}.

Limits also exist from rare meson decays \cite{Essig:2010gu} and dedicated solar axion experiments which rely only upon nuclear couplings \cite{Moriyama:1995bz, Krcmar:1998xn, Krcmar:2001si,Namba:2007rm, Derbin:2009yb, Belli:2012zz}, however as these are model-dependent we will not consider them going forward.

As we will demonstrate, there is also an entirely novel and model-independent constraint arising from axions emitted during nuclear transitions inside the Sun, which can be detected on Earth through the M1 process
\begin{equation}
	a+d\to n+p\,.
	\label{eq: axiodissociation}
\end{equation} 
where $d$ is a deuteron.
A particularly opportune target for this mechanism is the now-concluded Sudbury Neutrino Observatory (SNO) experiment, which relied upon a large quantity of deuterium to resolve the solar neutrino problem \cite{Boger:1999bb}.
The possibility of using deuterium for axion detection was first noted by Weinberg in \cite{Weinberg:1977ma}.

After detailing the corresponding axion flux and interaction cross section for \eqref{eq: axiodissociation} next, we then compute the resulting axion-induced event rate in SNO and derive constraints therefrom.
Conclusions and discussion are presented in closing.

\textbf{Solar axion flux.}
Whilst axions can in general be produced within stars by a number of processes, such as Primakoff conversion, electron bremsstrahlung and Compton scattering, we are primarily interested here in their production via low-lying nuclear transitions.
This is primarily because the threshold for deuterium dissociation via \eqref{eq: axiodissociation} is 2.2 MeV, and so only solar axions arising from nuclear transitions will have sufficient energy.
However there is also secondary benefit in so doing in that both emission and detection will rely only upon a single axion-nucleon coupling, allowing a model-independent constraint of general applicability.

Of the nuclear transitions occurring inside the sun, the most intense axion flux is provided by 
\begin{equation}
	p + d \to {}^3 \mathrm{He}+\gamma\left(5.5\mathrm{MeV}\right)\,,
	\label{eq: source process}
\end{equation}
where an axion substitutes for the emitted gamma ray.
In the Standard Solar Model (SSM) this constitutes the second stage of the $pp$ solar fusion chain, with the first stage provided by the two reactions $p+p\to d+e^++\nu_e$ and $p+p+e^-\to d+\nu_e$.
As the deuterons produced via this first stage capture protons within $\tau \simeq 6$ s, the axion flux resulting from \eqref{eq: source process} can be expressed in terms of the known $pp$ neutrino flux.

The constant of proportionality is the probability for a given M1 nuclear transition to result in axion rather than photon emission,
\begin{equation}
	\frac{\Gamma_a}{\Gamma_\gamma}\simeq\frac{1}{2\pi\alpha}\frac{m_n^2}{1+\delta^2}\left(\frac{\beta g_{aN}^1+g_{aN}^3}{\left(\mu_0-0.5\right)\beta+\mu_3-\eta}\right)^2\left(\frac{p_a}{p_\gamma}\right)^3\,,
	\label{eq:branching fraction}
\end{equation}
where $\alpha$ is the fine structure constant, $\delta^2=E/M$ is the relative probability for $E$ and $M$ transitions, we set $m_n=m_p$, and $\mu_0= \mu_p+\mu_n\simeq 0.88$ and $\mu_3=\mu_p-\mu_n\simeq 4.71$ are respectively the isoscalar and isovector nuclear magnetic moments, and $p_{a/\gamma}$ are the axion/photon momenta \cite{Avignone:1988bv}.
Dependence on specific nuclear matrix elements enters through $\beta$ and $\eta$.

In \eqref{eq: source process}, the M1-type transitions associated to axion emission correspond to capture of protons with zero orbital momentum. 
The probability of this occurring at a proton energy of 1 keV has been measured and found to be 0.55 \cite{Schmid:1997zz}, implying $\delta^2=0.82$. 
Since capture from the S state corresponds to an isovector transition, we can assume the $\beta g_{aN}^1$ contribution to \eqref{eq:branching fraction} is negligible and, to a good approximation, also ignore everything other than $\mu_3$ in the denominator \cite{Donnelly:1978ty}, leading to
\begin{equation}
	\frac{\Gamma_a}{\Gamma_\gamma}\simeq\frac{1}{2\pi\alpha}\frac{m_n^2}{1+\delta^2}\left(\frac{g_{aN}^3}{\mu_3}\right)^2\left(\frac{p_a}{p_\gamma}\right)^3\,.
\end{equation}
Following \cite{Bellini:2012kz} this provides the flux at Earth due to \eqref{eq: source process},
\begin{equation}
	\phi_a\simeq3.23\times 10^{10} m_n^2(g_{aN}^3)^2\left({p_a}/{p_\gamma}\right)^3\mathrm{cm}^{-2}\mathrm{s}^{-1}\,. 
	\label{eq: source flux}
\end{equation}

As this component of the solar axion flux has already been well explored by the Borexino and CAST experiments via a number of detection channels \cite{Bellini:2012kz, Andriamonje:2009ar}, we will in the following focus only on the previously unexplored detection channel provided by deuterium \eqref{eq: axiodissociation}.

\textbf{Axiodissociation cross section}
We now construct the cross section for the `axiodissociation' process
\begin{equation}
	a+d\to n+p\,,
	\label{eq:axiodissociation process}
\end{equation}
with threshold energy 2.2 MeV.
For reasons of clarity and brevity we give a relatively simple derivation of this quantity, postponing more thorough study to the future.
Details of the corresponding nuclear physics are found in \cite{Sachs:1953,Segre:1965}.

Working in the rest frame of the initial state deuteron, the number of events we expect is
\begin{equation}
	N_e=\sigma\phi N_dT\,,
	\label{event number}
\end{equation}
where $\phi$ is the incoming flux, $N_d$ the number of targets and $T$ the time, we can rearrange for a single deuteron to give $\sigma=(V/v_i)(N_e/T)$, where we have used the fact that for a single incoming axion $\phi\equiv n_iv_i=v_i/V$.
No integration over energy is required since thermal broadening is negligible relative to the axion energy at hand.

Identifying $N_e/T$ as the transition rate per unit time, we can make use of Fermi's golden rule to write
\begin{equation}
	\sigma=\frac{2\pi V}{v_i} |\langle f| H_I |i\rangle |^2\rho(k)\,, \quad \rho(k)=\frac{dn}{dE_k}\,,
	\label{cross section}
\end{equation}
where the initial and final states are 
\begin{equation}
	| i \rangle=|{}^3S{}_1; q\rangle\,,\quad
	| f \rangle=|{}^1S{}_0; 0\rangle\,,
\end{equation}
a deuteron in the ${}^3$S${}_1$ ground state and an incoming axion, and the continuum ${}^1$S${}_0$ state, while \eqref{derivative action} gives the non-relativistic interaction Hamiltonian
\begin{equation}
	H_I\simeq\frac{1}{2}\left(g^1_{aN}I+ g^3_{aN}\tau_3\right)\vec\nabla a\cdot\vec\sigma\,.
\end{equation}

We can expand in momentum modes via
\begin{equation}
	a(x)=\frac{1}{\sqrt{V}}\sum_{q'}\frac{1}{\sqrt{2E_{q'}}}\left(a(q')e^{iq'\cdot x}+a^\dagger(q')e^{-iq'\cdot x}\right)\,,
\end{equation}
so that due to the raising operation $\langle 0|a(q')=\langle q'|$, the axion part of the matrix element contains $\langle 0| a(q')| q \rangle = \delta_{qq'}$, which removes the summation over modes. 
Taking the derivative to get a factor of $\vec q$, we then have
\begin{equation}
	\langle f| H_I |i\rangle = \frac{i}{\sqrt{8E_qV}}\langle {}^1S{}_0|\left(g^1_{aN}I+g^3_{aN}\tau_3\right)\ \vec\sigma\cdot\vec q e^{iq\cdot x} | {}^3S{}_1\rangle\,,
\end{equation}
where the exponential factor can be neglected as a long-wavelength approximation.
To account for the isospin structure we can rewrite 
\begin{equation}
	 \left(g^1_{aN}I+g^3_{aN}\tau_3\right)\vec\sigma= g^1_{aN}(\vec\sigma_n+\vec\sigma_p)+g^3_{aN}(\vec\sigma_n-\vec\sigma_p)\,,
\end{equation}
where the first term gives zero acting on the outgoing singlet state, since in that case the neutron and proton spins are anti-aligned, so that we then have
\begin{equation}
	\langle f| H_I |i\rangle = \frac{i|\vec q|g^3_{aN}}{\sqrt{8E_qV}}\langle {}^1S{}_0|\hat q\cdot(\vec\sigma_n-\vec\sigma_p) | {}^3S{}_1\rangle\,.
\end{equation}

To evaluate this expression fully, we need to solve the Schr\"odinger equation for the deuteron wavefunction.
Although in principle a two-body problem, we can reduce this to a one-body problem satisfying
\begin{equation}
	\frac{1}{2m}\frac{1}{r}\frac{\partial^2}{\partial r^2}(r\psi(r))+(E_D-V(r))\psi(r)=0\,,\quad
\end{equation}
where $m\simeq m_n/2$ is the reduced mass, $r$ is the distance between nucleons, $E_D= 2.2$ MeV the binding energy of the deuteron and $\psi$ and $V$ are functions of $r$ only due to the spherical symmetry of the presumed S-wave ground state.
For the simplest possible approximation we can take $V_0$ corresponding to a delta function potential at the origin, which then yields 
\begin{equation}
	\psi_3=\frac{1}{\sqrt{4\pi r^2}}u(r)\chi_3\,,\quad 
	u(r)=Ne^{-\alpha r}\,,
\end{equation}
where $N=\sqrt{2\alpha}$ and  $\alpha = \sqrt{m_n E_D}$, which satisfies the normalisation condition $\langle\psi |\psi \rangle=\int \psi^*\psi \,dV=1$.
The spin of the triplet is encoded in the eigenfunction
\begin{align}
	\chi_3^m=
	\begin{cases} 
		|n^\uparrow p^\uparrow\rangle& S_z=+1\\ 
		\frac{1}{\sqrt{2}}(|n^\uparrow p^\downarrow\rangle +|n^\downarrow p^\uparrow \rangle)& S_z=0\\
		|n^\downarrow p^\downarrow\rangle & S_z=-1
	\end{cases}\,,
\end{align}
where $S_z$ is the $z$ component of the spin.

The eigenfunction of the outgoing singlet is 
\begin{equation}
	\psi_0=\frac{1}{\sqrt{4\pi r^2}}j(r)\chi_0\,,
	\quad j(r)=\sqrt{\frac{2}{L}}\sin(kr+\delta_0)\,,
\end{equation}
where $\delta_0$ is the s-wave phase shift produced by the singlet potential and $\chi_0=\frac{1}{\sqrt{2}}(|n^\uparrow p^\downarrow\rangle -|n^\downarrow p^\uparrow \rangle)$.
With the boundary condition $kL+\delta_0=n\pi$ it is straightforward to check that $\int |\psi_0|^2dV \to 1$ as $L\to\infty$.
We can then write 
\begin{equation}	
	\langle {}^1S{}_0|\hat q\cdot(\vec\sigma_n-\vec\sigma_p)| {}^3S{}_1\rangle
	=\int j(r)^*u(r) (\chi_0, \hat q\cdot(\vec\sigma_n-\vec\sigma_p)\chi_3) \,dr\,,
\end{equation}
where $(\psi,\chi)$ denotes the usual inner product for spinors.
For the $r$-dependent piece we have
\begin{equation}
	\int j(r)^*u(r) \,dr
	=\sqrt{\frac{2}{L}}\frac{N(\alpha\sin(\delta_0) +k\cos(\delta_0))}{k^2+\alpha^2}\,.
\end{equation}
We can further introduce the singlet scattering length $a_s = -1/k \cot(\delta_0)=-23.7$ fm, to yield
\begin{equation}
	\int j(r)^*u(r) \,dr =\sqrt{\frac{2}{L}} \frac{Nk(1-\alpha a_s)}{(k^2+\alpha^2)\sqrt{1+k^2a_s^2}}\,.
\end{equation}

The spin-dependent factor in $|\langle f| H_I |i\rangle|^2$, given that we must average over all the incoming deuteron polarisations, is $\frac{1}{3}\sum_m|(\chi_0, e\cdot(\vec\sigma_n-\vec\sigma_p)\chi_3^m)|^2$, where we sum over all possible spin states of the triplet ground state.
Considering the $z$ component of $\vec\sigma$ acting on the spin-dependent part of the wavefunction, $\sigma^z|\uparrow\rangle=|\uparrow\rangle$ and $\sigma^z|\downarrow\rangle=-|\downarrow\rangle$ so that for the singlet state of the deuteron we have
\begin{align}
	(\sigma_n-\sigma_p)^z \frac{1}{\sqrt{2}}(|n^\uparrow p^\downarrow\rangle -|n^\downarrow p^\uparrow \rangle)
	=\frac{2}{\sqrt{2}}(|n^\uparrow p^\downarrow\rangle +|n^\downarrow p^\uparrow \rangle)\,,
\end{align}
which is twice the $S_z=0$ eigenfunction for the incoming ${}^3S_1$ state.
As this implies $(\chi_0, (\vec\sigma_n-\vec\sigma_p)\chi_0)=0$ we can extend the sum to cover all possible states and then rewrite the spin-dependent factor as
\begin{equation}
	\frac{1}{3}\sum_m (\chi_0, \hat q\cdot(\vec\sigma_n-\vec\sigma_p)\chi^m)(\chi^m, \hat q\cdot(\vec\sigma_n-\vec\sigma_p)\chi_0)\,.
\end{equation}
Recognising the insertion of $\sum_m |\chi^m\rangle\langle\chi^m|=1$, this is
\begin{equation}
	\frac{1}{3}(\chi_0, (\hat q\cdot(\vec\sigma_n-\vec\sigma_p))^2 \chi_0)=\frac{4}{3}\,,
\end{equation}
where we have used $\vec\sigma_n\chi_0=-\vec\sigma_p\chi_0$ and $\sigma^{(i}\sigma^{j)}=\delta^{ij}$.

From the boundary condition $kL+\delta_0=n\pi$ we have a single state for each $n$, so that the number of outgoing states between $k$ and $k+dk$ is 
\begin{equation}
	dn = \frac{1}{\pi}\left(L+\frac{d\delta_0}{dk}\right)dk\big|_{L\to\infty}=\frac{L}{\pi}dk\,.
\end{equation}
Given that $d k/dE_k=1/v_f$, from \eqref{cross section} we then have
\begin{equation}
	\sigma=\frac{2V L}{v_iv_f}|\langle f| H_I |i\rangle |^2\,.
\end{equation}
Using $\vec qc=\gamma m_0 c^2 v_i = E_a v_i$ and $|\vec k|= \sqrt{m_n(E_a-E_D)}$,
\begin{equation}
	\sigma
	=\frac{2}{3}(g^3_{aN})^2 m_n\sqrt{E_a^2-m_a^2}\frac{|\vec k|\alpha}{(k^2+\alpha^2)^2}\frac{(1-\alpha a_s)^2}{1+k^2a_s^2}\,.
\end{equation}

\textbf{Data analysis.}
Given the axion flux and cross section provided previously, we can then calculate the axion-induced event rate in the detector via \eqref{event number}.
It is of course important to emphasise that the original purpose of the deuterium in SNO was to observe the neutral current (NC) process
\begin{equation}
	\nu_x+d\to n + p + \nu_x\,,
	\label{neutrino dissociation}
\end{equation}
which could then, in concert with measurements of the electron neutrino flux, conclusively betray the presence of solar neutrino flavour oscillations.
As the final state neutrino is invisible to the detector, the signature of this process is at first sight identical to that of \eqref{eq:axiodissociation process}.

Of course one key difference exists in that the 5.5 MeV axions produced via \eqref{eq: source process} are monoenergetic, whilst the ${}^8B$ neutrinos driving \eqref{neutrino dissociation} have a continuous energy spectrum with an endpoint near 15.8 MeV.
However these spectral differences are largely washed out by the process of neutron thermalisation, which negates sensitivity to the spectral character of the input flux.

More specifically, detection of either of these phenomena rests upon the liberated neutron, once sufficiently thermalised, being recaptured. 
This can occur on a deuteron, resulting the emission of a 6.25 MeV gamma ray, or alternatively in the phase II SNO dataset onto a ${}^{35}$Cl nucleus, resulting the emission of 8.6 MeV in gamma rays.
For the phase III dataset the ${}^{35}$Cl was removed and dedicated neutral current detectors (NCD) were introduced to enable non-Cherenkov detection of NC neutrons, via their capture onto ${}^3$He and the subsequent emission of a proton/triton pair carrying 0.764 MeV of energy \cite{Aharmim:2011vm}.
Since neutron capture cross sections are suppressed by their velocity, these processes will in general only occur once most of the initial kinetic energy has been lost, rendering SNO likely unable to distinguish between axion and neutrino-induced dissociated events.

Whilst this apparent loss of spectral information seems to remove the possibility of leveraging the monoenergetic nature of the axion flux to gain greater sensitivity, it does nonetheless simplify the required data analysis.

Following the approach employed in \cite{Grifols:2004yn}, we can write the total number of dissociation events seen by SNO as 
\begin{equation}
	N^{exp}=N_a+N_dT\phi_{SSM}\int \,dE_\nu\, \xi(E_\nu)\sigma_{NC}\,,
\end{equation}
where $\phi_{SSM}=(5.60\pm0.66)\times10^6$ cm${}^{-2}$ s${}^{-1}$ is the predicted value of the ${}^8$B solar neutrino flux in the B16-GS98 Standard Solar Model (SSM) \cite{Vinyoles:2017}, with spectral shape parametrised via
\begin{equation}
	\xi(E_\nu)=8.52\times 10^{-6}\left(15.1-\frac{E_\nu}{\mathrm{MeV}}\right)^{2.75}\left(\frac{E_\nu}{\mathrm{MeV}}\right)^2\,,
\end{equation}
and $\sigma_{NC}$ the neutral current neutrino-deuteron interaction cross section \cite{Nakamura:2000vp}.
As the resulting flux inferred by SNO assumes that these events arise only due to NC interactions, 
\begin{equation}
	\phi_{SNO}\equiv \frac{N^{exp}}{N_dT\int \,dE_\nu\, \xi(E_\nu)\sigma_{NC}}\,,
\end{equation}
and we can then write
\begin{equation}
	\phi_{SNO}=\phi_{SSM}+\frac{\phi_a \sigma_d }{\int \,dE_\nu\, \xi(E_\nu)\sigma_{NC}}\,,
\end{equation}
where no integration over energy is required since thermal broadening is negligible relative to the axion energy at hand.

Since axion and neutrino-induced dissociations cannot be distinguished by SNO, the former can lead to an excess of events over the SSM expectation, and the product $\phi_a \sigma_d$ is then constrained by the combined analysis of all three phases of SNO data, which yields $\phi_{SNO}=(5.25\pm0.20)\times10^6$ cm${}^{-2}$ s${}^{-1}$ \cite{Aharmim:2011vm}, where we have added errors in quadrature. 
Requiring that the $(\phi_{SNO}-\phi_{SSM})$ confidence interval not exceed 95 \% C.L. limits then provides the exclusion presented in Figure 1.

\begin{figure}
	\centering
	\includegraphics[width=1\columnwidth]{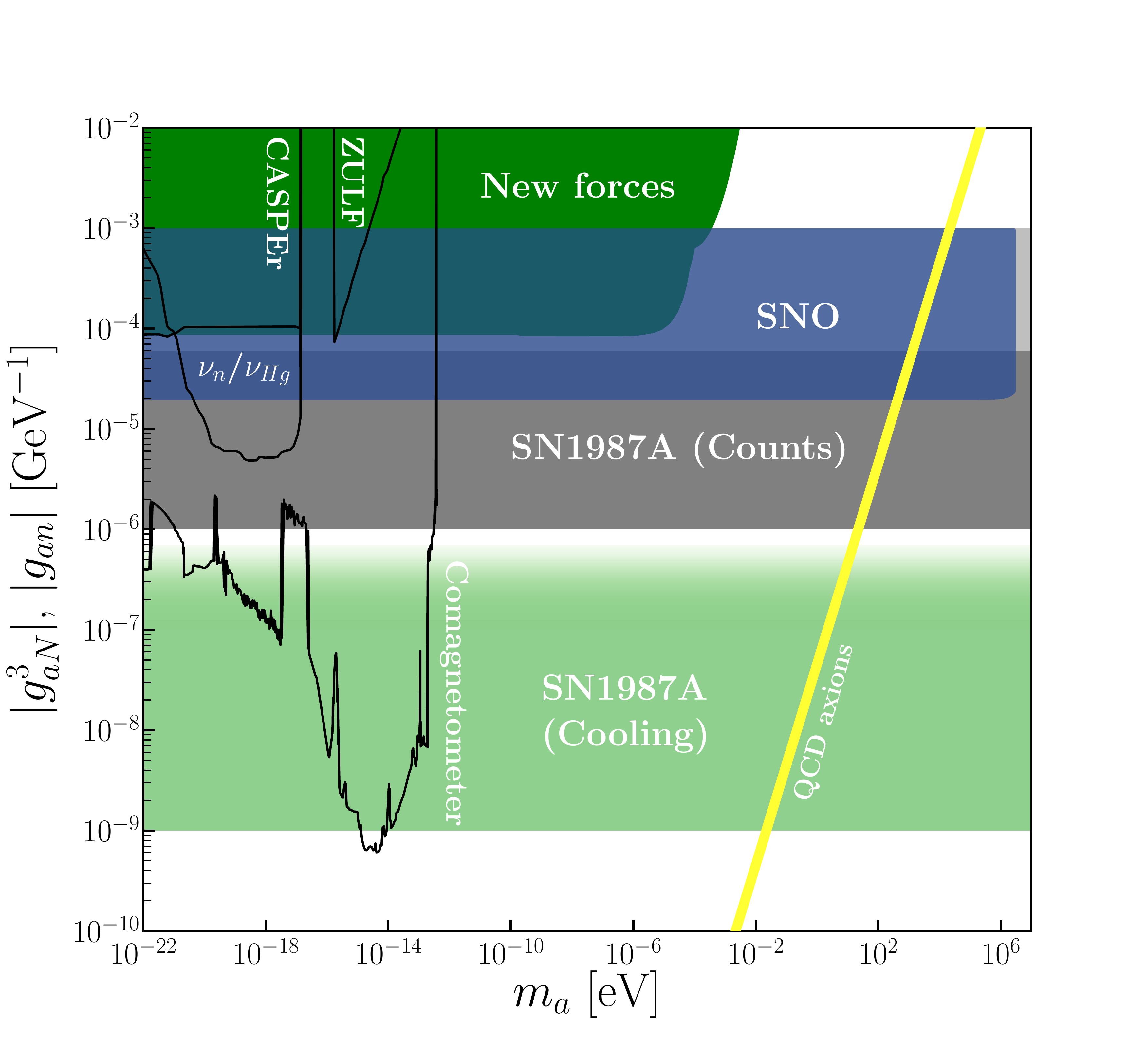}
	\caption{95 \% CL Constraint on the isovector axion-nucleon coupling $g_{aN}^3$ arising from axion-induced dissociation of deuterium at SNO (blue). 
	Following \cite{Bellini:2012kz, Raffelt:1982dr} we have imposed a `solar trapping' upper bound of $|g_{aN}^3|\sim10^{-3}$GeV${}^{-1}$, where axions interact sufficiently strongly to be trapped within the sun and hence escape detection.
	Also shown is the $g_{aN}^3$-sensitive component of the SN1987A constraint from \cite{Engel:1990zd} (dark gray) arising from additional event counts at Kamiokande, even though this is not strictly comparable to our result due to dependence on couplings other than $g_{aN}^3$.
	In the absence of a precise cancellation our result also limits $g_{an}$ and $g_{ap}$ individually, so constraints on the former are provided as a visual reference.
	These include SN1987A cooling \cite{Carenza:2019pxu} (light green), SN1987A counts \cite{Engel:1990zd} (light grey), and the non-observation of new forces \cite{Vasilakis:2008yn,  Adelberger:2006dh} (dark green).
	Constraints which rely on the axion comprising all of the observed dark matter are also outlined (nuclear spin precession in cold neutrons/Hg \cite{Abel:2017rtm}, CASPER \cite{Wu:2019exd}, CASPER-ZULF \cite{Garcon:2019inh} and old comagnetometer data \cite{Bloch:2019lcy}).
	The QCD axion band for $g_{aN}^3$ is given in yellow.}
	\label{fig:results}
\end{figure}

As can be seen, we exclude $|g_{aN}^3|$ between $2\times10^{-5}$ and $10^{-3}$ GeV${}^{-1}$, for axion masses up to $\sim5.5$ MeV.
For larger couplings this improves upon the $g_{aN}^3$-sensitive component of the SN1987A constraint in Ref.~\cite{Engel:1990zd} arising from additional particle-emission counts at Kamiokande, even though this is not strictly comparable as the SN1987A axion flux is not solely dependent on $g_{aN}^3$. 

The QCD axion band bounded by the KSVZ and DFSZ models with $\cos^2\beta =1$ is also shown; the latter case we exclude for axion masses between 0.3 and 13 keV, although this region is in any case ruled out by astrophysical considerations and direct results from PandaX-II \cite{Fu:2017lfc}.
Variant QCD axion models may of course differ substantially from this benchmark \cite{Gao:2019tqt}.

Constraints on $g_{an}$ are also shown, if we assume no precise cancellation between $g_{an}$ and $g_{ap}$ then any limit on $g_{aN}^3$ is equivalent to a limit on both $g_{an}$ and $g_{ap}$. 
This assumption is not too restrictive; in the analysis of Ref.~\cite{DiLuzio:2017ogq} this cancellation requires a DFSZ-type model with the specific tuning $\tan\beta\simeq\sqrt{2}$ or $1/\sqrt{2}$.

This being the case our result can also exceed comparable constraints from other laboratory experiments, and exclude regions of the parameter space for which astrophysical constraints from SN1987A and NS cooling are inapplicable due to axion trapping, where axions cannot free stream and contribute to anomalous cooling as they do in other regions of the parameter space.
The latter constraints, which in any case share some degree of degeneracy with the SN1987A case, are not included in Figure 1 due to model-dependence and the unknown shape of the exclusion region for non-negligible axion masses.

\textbf{Conclusions and discussion.}
The axion is a notably well-motivated aspect of physics beyond the Standard Model, and as such has been a topic of much investigation in recent years.
Nonetheless, a large majority of these studies are based upon the interactions of axions with electromagnetism, leaving their couplings to nucleons and other species comparatively less well explored.

In this paper we have established a novel detection channel sensitive to precisely one of these couplings, relying upon the ability of suitably energetic axions to dissociate deuterons into their constituent neutrons and protons.
In concert with the 5.5 MeV solar axion flux arising from the $p + d \to {}^3 \mathrm{He}+a$ process, one can then derive a model-independent constraint on the isovector axion-nucleon coupling $g_{aN}^3$.
A particularly opportune target for this search strategy is the now-completed SNO experiment, which relied upon large quantities of deuterium to resolve the solar neutrino problem.

Having derived the corresponding `axiodissociation' cross section, from the full SNO dataset we exclude regions where $|g_{aN}^3|$ is between $2\times10^{-5}$ and $10^{-3}$ GeV${}^{-1}$, for axion masses up to $\sim5.5$ MeV, covering previously unexplored regions of the axion parameter space.
This comes with the added benefit that we do not require any assumptions about the nature of dark matter, or the astrophysics of SN1987A.

If furthermore we assume no precise cancellation between $g_{ap}$ and $g_{an}$ then any limit on $g_{aN}^3$ is equivalent to a limit on both $g_{an}$ and $g_{ap}$. 
In that case our result can exceed comparable constraints from other laboratory experiments, and exclude regions of the parameter space for which astrophysical constraints from SN1987A and NS cooling are inapplicable due to axion trapping.
Constraints on $g_{an}/g_{ap}$ from SN1987A event counts can probe equally large couplings, but at the cost of a variety of associated assumptions and uncertainties related to the modelling of SN1987A.

These findings could conceivably be improved in the future via a more sophisticated estimation of the axiodissociation cross section; we have neglected relativistic and finite-size corrections, along with D-state effects in the deuteron ground state.
The existence of other axion search strategies which rely only upon nuclear couplings is also a topic of ongoing investigation.

\begin{acknowledgments}
This work is supported by the National Key Research and Development Program of China Grant No. 2020YFC2201504, by the Projects No. 11875062 and No. 11947302 supported by the National Natural Science Foundation of China, and by the Key Research Program of Frontier Science, CAS. 
A. B. was supported by a Chinese Government Scholarship, and N.H. also acknowledges the hospitality of the Center for High Energy Physics at Peking University, where much of this work was completed.
\end{acknowledgments}

\end{document}